\documentclass[superscriptaddress,aps,prl,twocolumn,showpacs,preprintnumbers,amsmath,amssymb,floatfix,groupedaddress]{revtex4}
 
\usepackage{graphicx}
\usepackage{dcolumn}
\usepackage{bm}
 
\def\e{\mathcal{E}}
 
\begin{document}
 
\title{Realization of Coherent Optically Dense Media via Buffer-Gas Cooling} 
 
\author{Tao~Hong}
\affiliation{Harvard-MIT Center for Ultracold Atoms, Department of Physics, Harvard University, 17 Oxford Street, Cambridge, MA 02138}
\affiliation{Joint Quantum Institute, Department of Physics, University of Maryland, College Park, MD 20742}
\author{Alexey~V.~Gorshkov}
\author{David~Patterson}
\author{Alexander~S.~Zibrov} 
\author{John~M.~Doyle} 
\author{Mikhail~D.~Lukin} 
\author{Mara~G.~Prentiss}
\affiliation{Harvard-MIT Center for Ultracold Atoms, Department of Physics, Harvard University, 17 Oxford Street, Cambridge, MA 02138} 
 
\date{\today}
 
\begin{abstract}
 
We demonstrate that buffer-gas cooling combined with laser ablation can be used to create  
coherent optical media with high optical depth and low Doppler broadening that offers metastable states with low collisional and motional decoherence. Demonstration of this generic technique opens pathways to coherent optics with a large variety of atoms and molecules.  
We use helium buffer gas to cool $^{87}$Rb atoms to below 7 K and slow atom diffusion to the walls. Electromagnetically induced transparency (EIT) in this medium allows for 50\% transmission in a 
medium with initial OD $>70$ and for slow pulse propagation with large delay-bandwidth products. 
In the high-OD regime, we observe high-contrast spectrum oscillations due to efficient four-wave mixing.

\end{abstract}
 
 
 
 
\maketitle
 
Greatly facilitated by the advent of electromagnetically induced transparency (EIT) \cite{harris97, fleischhauer05}, coherent optical atomic media have been widely used in quantum and nonlinear optics \cite{hemmer95harris97b,imamoglu97harris98hong03}, as well as in high precision measurements  \cite{kominis03hong05kornack05}. For example, they underlie few-photon nonlinear optics \cite{imamoglu97harris98hong03} and resonant enhancement of 
wave mixing \cite{hemmer95harris97b}. 
In addition, motivated by ideas from quantum information science,  EIT is 
now widely used  \cite{eisaman05, appel08, chaneliere05, choi08, honda08,  kolchin06, yuan07, simon07} in  applications to  quantum memories \cite{lukin03, fleischhauer} and to long-distance quantum communication \cite{duan01}. 
Practical use of 
these applications 
would be greatly facilitated by more scalable simpler technologies that can simultaneously achieve high optical depth (OD) and long lifetimes for metastable 
state coherences. For example, in quantum memories, OD 
and decoherence rates determine efficiency and memory time, respectively \cite{lukin03}.  However, the two currently available atomic vapor systems - hot atoms in vapor cells \cite{eisaman05, appel08} and cold atoms in magneto-optical traps (MOTs) \cite{chaneliere05, choi08, honda08, kolchin06, yuan07} - typically operate at OD $\lesssim 10$ and suffer from relatively fast decoherence of hyperfine sublevels. Although recently several novel experimental 
systems, such as high-aspect-ratio MOTs  \cite{vengalattore05harris08} and cavity-enhanced MOTs  \cite{simon07}, have been shown to exhibit superior performance, improving the scalability of these systems and extending them to other atomic or molecular species 
is highly nontrivial due to 
restrictions imposed by laser cooling. 

In this Letter, we integrate 
buffer-gas cooling and laser ablation to demonstrate a new and simple approach to producing coherent optically dense media. Using Rb as an example, we 
achieve 
high OD ($> 70$) combined with a low ground state decoherence rate by cooling Rb vapor to below 7 K using a He buffer gas. This system combines the fundamental advantages of MOT systems (slow atoms with small Doppler broadening) with large volumes and simplicity characteristic of hot vapor cells. 
This out-of-equilibrium thermal system offers high density metallic gases at temperatures hundreds or thousands of times colder than the boiling point of the metal. 
Our work is the first application of buffer-gas-cooled media to coherent quantum optics. 
In particular, 
EIT 
increases peak transmission in our system by a factor of $\sim \!\! \exp(70)$, while 
maintaining ground hyperfine decoherence rates on the order of 10  kHz. We also 
observe parametric four-wave mixing processes \cite{mikhailov03boyer07kang04wong04harada04agarwal06}  and the resultant splitting of the EIT transmission 
peak into several peaks. Finally, we demonstrate slow pulse propagation with large delay-bandwidth products.

\begin{figure}[b]
\includegraphics[scale=0.6]{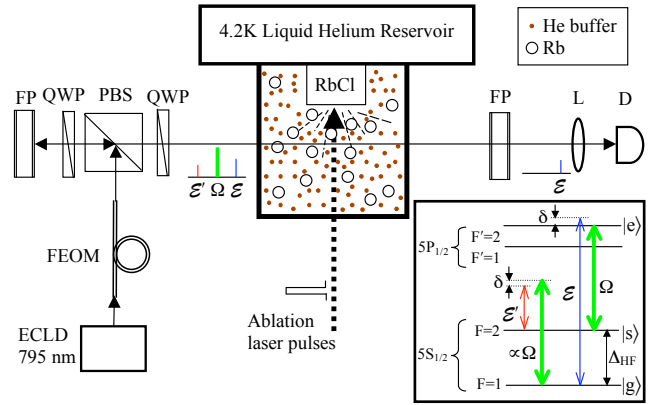}
\caption{\label{fig:apparatus} (Color online) Experimental setup (PBS, 
polarizing beam splitter; QWP, quarter-wave plate; L, lens; D, photodiode 
detector). The inset shows the $^{87}$Rb level structure.}
\end{figure}

\textit{Experimental Setup.--} 
Our cryostat system used liquid He to cool to 4.2 K a 2cm$\times$2cm$\times$2cm 
Al cell filled with He buffer gas and surrounded by Helmholtz coils to cancel 
stray magnetic fields (Fig.\ \ref{fig:apparatus}). Inside the cell, a compressed 
RbCl pillet was mounted as a target for laser ablation. It yielded Rb vapor when 
bombarded by a 532 nm pulsed YAG laser. Ablation pulses with typical energies of 
10 mJ and typical durations of 3-5 ns were used. Once generated, the hot Rb 
vapor quickly expanded until it filled the whole cell. After the ablation pulse, 
collisions with He cooled the Rb down to below 7 K in less than 1 ms. Typically,  
40 mTorr of cold He buffer gas was used (as measured by a room temperature 
pressure gauge). We note that the entire cryostat system is compact (30 cm tall and 20 cm diameter round), reliable and inexpensive to operate, using about 10 liters of liquid helium a day, including cooling the unit down from 300 K to 4 K. We study EIT on the Rb D1 line ($\lambda =$ 795 nm) by phase 
modulating the light from an external-cavity diode laser (ECLD) using a fiber 
electro-optical modulator (FEOM) operating at the ground state hyperfine 
splitting of $^{87}$Rb $\Delta_\textrm{HF} \approx (2 \pi) 6.8$ GHz.  As shown 
in the inset of Fig. \ref{fig:apparatus}, the laser carrier frequency that is 
resonant with the $F = 2 \rightarrow F' = 2$ transition, played the role of the 
control field with Rabi frequency $\Omega$. 
The high-frequency 
modulation sideband, tuned near the $F = 1 \rightarrow F' = 2$ transition with 
two-photon detuning $\delta$, played the role of the signal with amplitude $\e$. 
A Fabry-Perot (FP) etalon was used to block the control and the low-frequency 
off-resonant sideband $\e'$ after the cell. For the EIT observation  (but not 
for the four-wave mixing and slow light measurements), a second FP etalon before 
the cell was used to attenuate the off-resonant sideband intensity by a factor 
of $\approx 1/5$. All three optical fields were $\sigma^+$-polarized.  
The physics was dominated by the $\Lambda$-system formed by $|g\rangle = |F = 1, 
m = 1\rangle$, $|s\rangle = |F = 2, m = 1\rangle$, and $|e\rangle = |F' = 2, m = 
2\rangle$, which we used for modeling.   

\begin{figure}[tb]
\includegraphics[scale=0.30]{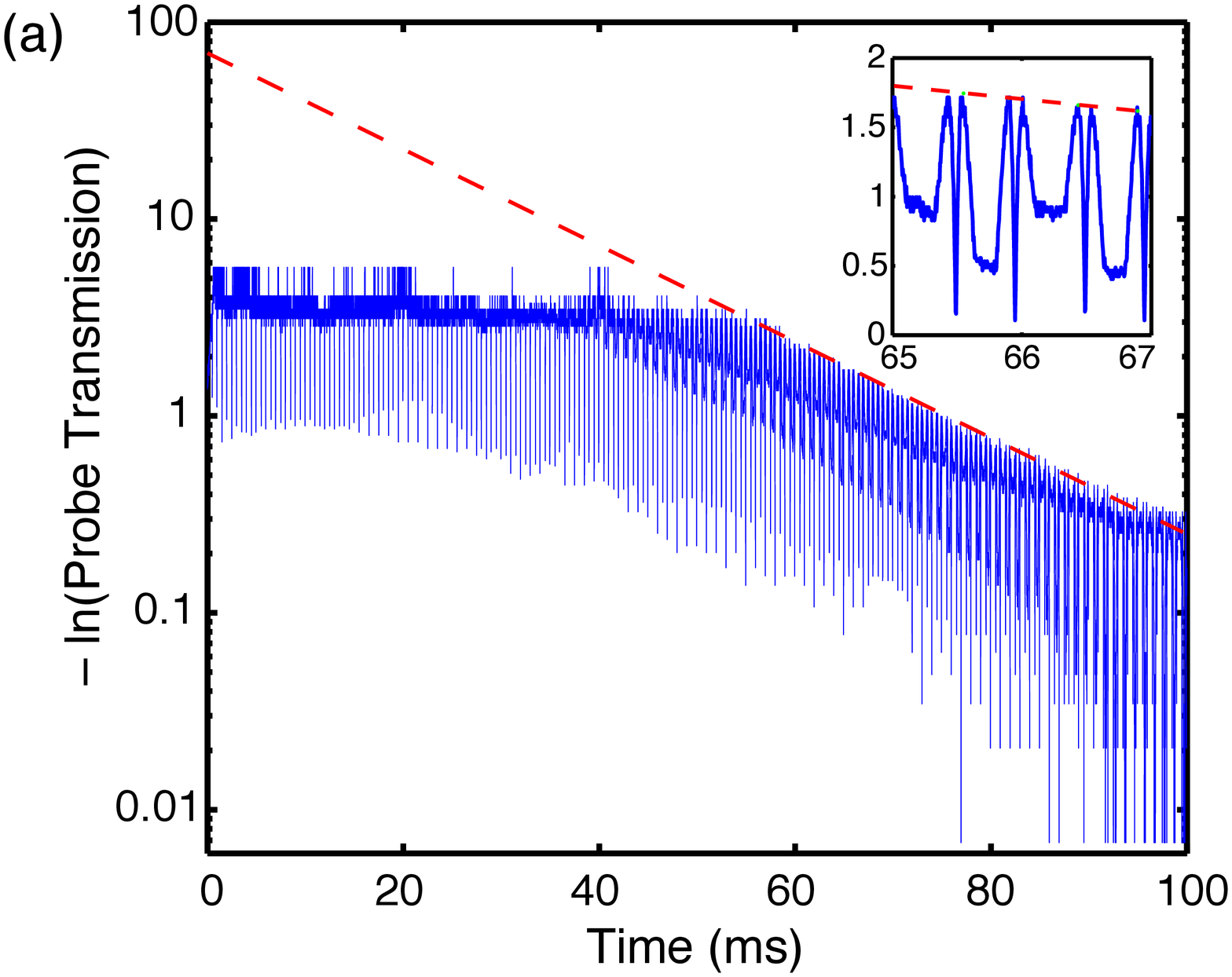}
\includegraphics[scale=0.40]{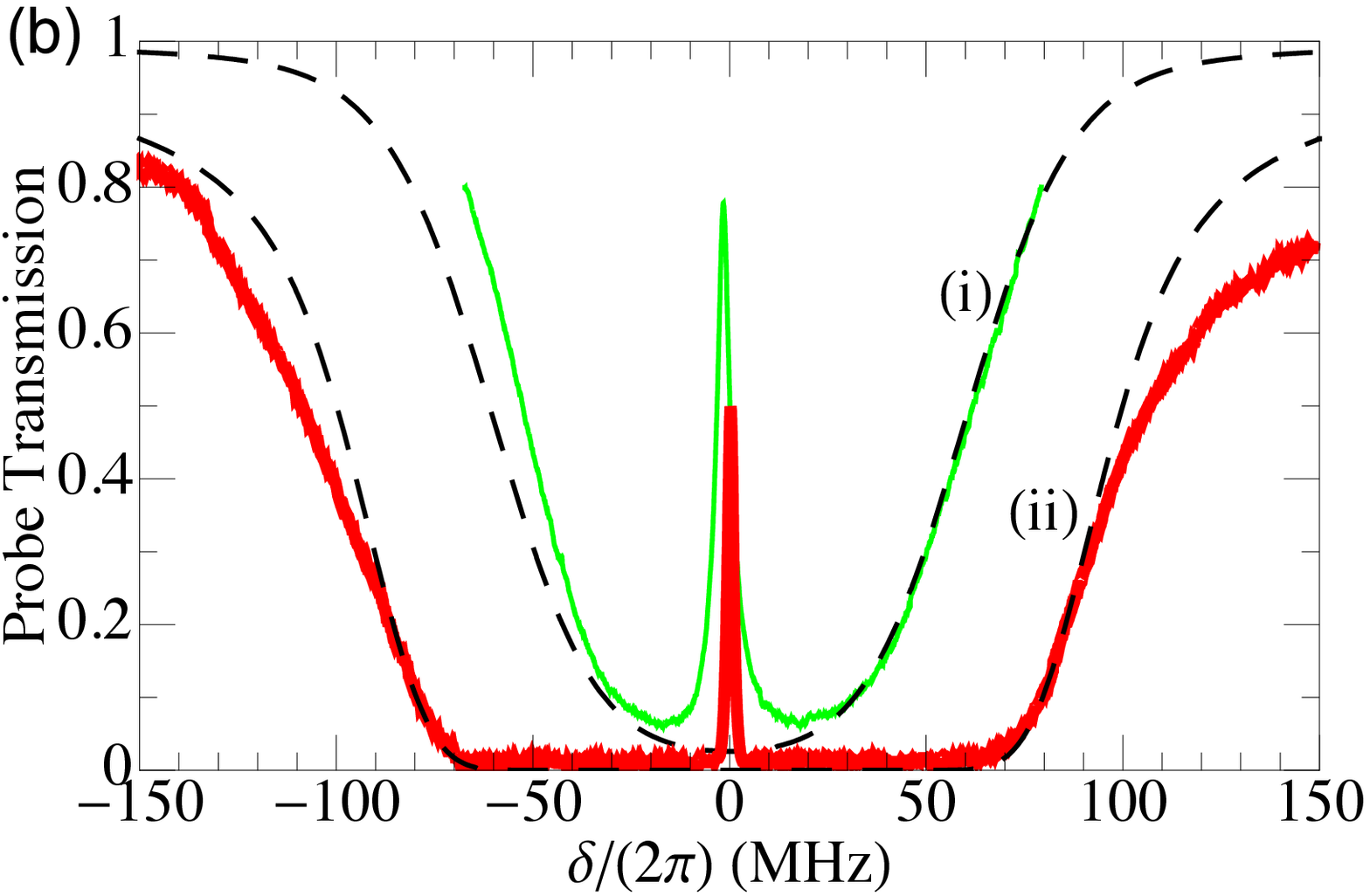}
\caption{(Color online) OD estimates in the presence of EIT.  (a) Solid blue is 
the negative logarithm of probe transmission as a function of time, as the probe 
frequency is scanned back and fourth through two-photon resonance resulting in 
the sharp EIT dips.  Red dashed line is an exponential fit to the background 
optical depth. The inset zooms in on the plot around $t = 66$ ms.  (b) Probe 
transmission (solid lines) at two different ODs exhibiting an EIT peak on the 
background of a Doppler-broadened single-photon absorption dip. Dashed curves 
are theoretical fits to the single-photon absorption dip. \label{fig:od}}
\end{figure}

\textit{EIT in an optically thick medium.--} 
The optical depth (OD) is defined so that if we were to turn off the control 
field after optically pumping the atoms, the intensity of a vanishingly weak 
resonant probe would be attenuated by $\exp(-\textrm{OD})$ after passing through 
the medium. In the experiment, however, a continuous-wave (CW) control field (60 
$\mu$W, $\approx$0.4 mm diameter) was used to provide continuous optical 
pumping. 
At the same time, two-photon detuning $\delta$ of a weak probe  (0.5 $\mu$W, 
$\approx$0.4 mm diameter) was continuously scanned back and forth through two-
photon resonance ($\delta = 0$) at a rate of $(2 \pi)$300 MHz/ms using a 
triangle wave. The measured  
probe transmission, thus, exhibited EIT peaks every 0.5 ms as a function of time 
while showing $\approx \exp(-\textrm{OD})$ in between the EIT peaks. To extract 
the OD, we plot in Fig.\ \ref{fig:od}(a) the negative logarithm of probe 
transmission as a function of time (solid blue line), in which EIT peaks show up 
as dips and OD (red dashed line) can be read out between the peaks, as shown in 
the inset.  After the ablation pulse (at time $t = 0$), diffusion of Rb through 
the He buffer gas and subsequent absorption of Rb onto the cold cell walls 
reduces the density of Rb [and hence OD$(t)$]. For $t < 50$ ms, OD$(t)$ is so 
large that the transmitted signal is completely immersed in the noise background 
making direct measurement of OD impossible. At larger $t$, however, OD$(t)$ 
becomes measurable and follows an exponential $\textrm{OD}(t) \approx 75 \exp[-
t/(17.5 \textrm{ms})]$ (red dashed line) \cite{odnote}. Extrapolating this fit 
toward $t = 0$, we obtain OD $\approx 75$ near $t = 0$. At these early times, 
EIT peaks show about 50\% transmission. To support the extrapolation of the 
fit to $t < 50$ ms, we performed an analogous measurement on $^{85}$Rb ($2\%$ 
abundance in our sample) and observed exponential behavior for $t < 50$ ms (see 
also Ref.\ \cite{sushkov08}).

In Fig.\ \ref{fig:od}(b), we give an independent demonstration that high optical 
depths were indeed obtained and exhibit typical narrow and tall EIT two-photon 
resonances. The two solid lines show probe transmission spectrum at two 
different ODs for the same control and probe intensities as in Fig.\ 
\ref{fig:od}(a). The single-photon Doppler broadened resonances were fit to the 
negative exponential  (dashed curve) of a Voigt profile made up from a 
convolution of a Gaussian with FWHM of $\Delta_\textrm{D}$ and a Lorentzian with 
natural FWHM of $2 \gamma = (2 \pi)5.7$ MHz. For the low-OD curve (i), 
$\Delta_\textrm{D}$ and OD were varied to find OD $ \approx 3.7$ and 
$\Delta_\textrm{D} = (2 \pi)75$ MHz (corresponding to an effective temperature 
of $7$ K).  
Taking this $\Delta_\textrm{D}$, which we will also use for the rest of the 
paper, and adjusting OD to match the flat part of the high-OD spectrum (ii), we 
obtain OD $ \approx 35$. 
We also note a narrow [$(2 \pi) 1.8$ MHz] and 
high-contrast (50\% transmission) EIT two-photon resonance observed in the high-
OD curve (ii). 

\begin{figure}[tb]
\includegraphics[scale=0.37]{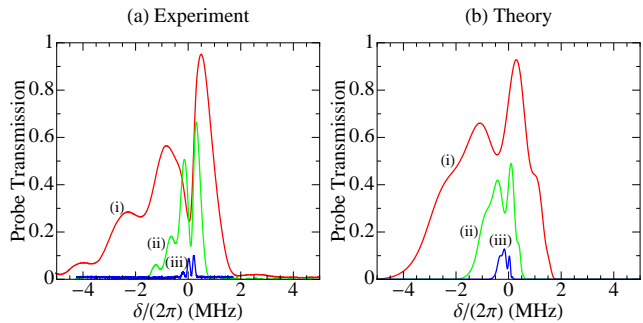}
\caption{(Color online) Interference fringes in the EIT spectrum caused by four-
wave mixing. Control field powers are (i) 0.58 mW, (ii) 0.21 mW, and (iii) 0.077 
mW, with beam diameter of 0.4 mm (the latter deduced from the Rabi frequency 
fit). (a) Experiment. Since the absolute value of $\delta$ was not calibrated 
experimentally, $\delta = 0$ is chosen to approximately match the theory. (b) 
Theory. \label{fig:fourwave} }
\end{figure}

\textit{Four-wave mixing.--}  At high optical depths, the observed EIT 
transmission peak exhibits oscillations whose number increases as the optical 
depth increases. In Fig.\ \ref{fig:fourwave}(a), we show probe transmission for 
a fixed OD and for different control powers. As explained below, the observed 
oscillations correspond to interference fringes between the directly transmitted 
probe output and a contribution due to four-wave mixing. Modulation is caused by 
the dispersion associated with EIT.

To analyze such effects quantitatively, we include the effects of four-wave 
mixing by following the analysis in Refs.\ \cite{lukin97, lukin98b, lukin99}. 
The use of Floquet theory to include the far-off-resonant couplings to order 
$1/\Delta_\textrm{HF}$ yields an effective Rabi frequency $\propto 
1/\Delta_\textrm{HF}$ coupling $|g\rangle$ and $|s\rangle$.  We use the master 
equation to compute the polarizations driving $\e$ and $\e'$ to linear order in 
$\e$ and $\e'$. Ignoring the small Stark shift of $|g\rangle$ 
$\propto|\Omega|^2/\Delta_\textrm{HF}$ and integrating over the single-photon 
Doppler shift $\Delta$, we obtain steady-state propagation equations for $\e$ 
and $\e'$: 
\begin{eqnarray} \label{eep}
&\partial_z \left[
\begin{array}{c}
\e(z,\delta) \\
\e'^*(z,\delta) \\ 
\end{array}
\right] 
= \frac{1}{L} M(\delta) 
\left[
\begin{array}{c}
\e(z,\delta) \\
\e'^*(z,\delta) \\ 
\end{array}
\right],& \\ 
\label{M}
&M(\delta) = \int_{-\infty}^\infty d \Delta \frac{e^{-\frac{(\Delta-
\Delta_c)^2}{2 \sigma^2}}}{\sqrt{2 \pi} \sigma}  i \frac{d \gamma}{F} \left[
\begin{array}{ccc}
\delta + i \gamma_0 & -\frac{\Omega^2}{\Delta_\textrm{HF}} \\
\frac{\Omega^{*2}}{\Delta_\textrm{HF}} & 0 \\ 
\end{array}
\right].& 
\end{eqnarray}
Here $F = |\Omega|^2 - (\delta+\Delta + i \gamma)(\delta+ i \gamma_0)$, $L$ 
is the medium length, $\gamma_0$ is the decay rate of the $|s\rangle-|g\rangle$ 
coherence, $\sigma = \Delta_\textrm{D}/(2 \sqrt{2 \ln 2})$ is the standard 
deviation of the Doppler profile, and $d = \frac{3}{8 \pi} \lambda^2 L 
\frac{N}{V}$ would have been half of the resonant optical depth if there were no 
Doppler broadening [i.e.\ $|e(L,0)|^2 = \exp(-2 d) |e(0,0)|^2$ for 
$\Delta_\textrm{D} \rightarrow 0$ and  $\Omega = 0$]. $N/V$ is the density of 
atoms in state $|g\rangle$. For our values of $\gamma$ and $\Delta_D$, $d = 4.6$ 
OD. We have also allowed for a possible deviation $\Delta_c$ of the control 
field from resonance. 

Given $\e(0,\delta) = - \e'(0,\delta) = 1$, we solve Eq.\ (\ref{eep}) for probe 
transmission $|\e(L,\delta)|^2$ by exponentiating $M(\delta)$. The result is 
shown in Fig.\ \ref{fig:fourwave}(b). For curve (i), $\Omega = (2 \pi) 15$ MHz 
and OD $ = 100$ were chosen to approximately reproduce the total peak width and 
the fringe period. For (ii) and (iii), $\Omega$ was computed from the known 
intensity ratios. $\gamma_0 = (2 \pi) 30$ kHz was adjusted to approximately 
reproduce the relative amplitudes of the three curves. $\Delta_c = (2 \pi) 40$ 
MHz was chosen to reproduce the left-right asymmetry of the data.

To get insight into the nature of the oscillations, we note that 
$M_{11}(\delta)$ describes EIT propagation, while the off-diagonal elements of 
$M$ are the parametric coupling coefficients giving 4-wave mixing. Large off-
diagonal matrix elements would result in exponential growth (gain). In the 
present experiment, however, the four-wave mixing terms are small due to the 
large $\Delta_\textrm{HF}$, so that $\exp[M(\delta)]$ can be expanded to first 
order in $1/\Delta_\textrm{HF}$ to give
\begin{equation} \label{expand}
|\e(L, \delta)|^2 \approx \left| e^{M_{11}(\delta)} + 
\frac{\Omega^2}{\Delta_\textrm{HF} (\delta + i \gamma_0)} 
\left(e^{M_{11}(\delta)} - 1\right)\right|^2. 
\end{equation}
Thus, the probe spectrum is an interference of EIT [$\exp(M_{11})$] and 4-wave 
mixing ($\propto \!\! 1/\Delta_\textrm{HF}$). Since the phase of the EIT term is 
$\approx \delta/(v_g/L)$ [$v_g = L \Omega^2/(\gamma d)$ is the group velocity], 
while the phase of the 4-wave mixing term is approximately $\delta$-independent 
(except for a sign change at $\delta = 0$), the fringes have period $2 \pi 
v_g/L$. Notice that it is the homogeneous optical depth $2 d$ (not OD) that 
determines $v_g$ \cite{kash99}. 

The demonstrated interference fringes confirm that the medium combines high 
optical depths and long coherence times because lower OD and higher $\gamma_0$ 
give theoretical spectra inconsistent with the data. In particular, smaller OD 
increases the fringe period, while larger $\gamma_0$ reduces transmission and 
increases the ratio of the amplitudes of the three curves. The mismatch between 
experimental and theoretical curves most likely arises from features not 
included in our simple model. In particular, velocity-changing coherence-
preserving  collisions alleviate the effects of Doppler broadening  
\cite{pack06} and may explain the high contrast of the experimentally observed 
fringes. 

\begin{figure}[tb]
\includegraphics[scale=0.58]{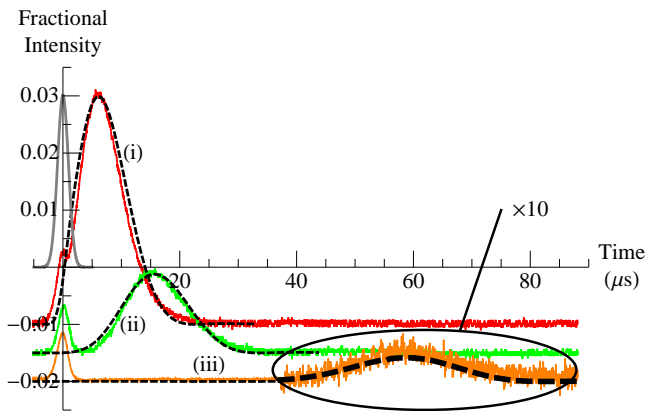}
\caption{(Color online) Pulse propagation through EIT medium at three different 
optical depths increasing from (i) to (iii). Intensity of 1 corresponds to the 
maximum intensity of the incident pulse, a down-scaled version of which (gray) 
is centered at time = 0. The three experimental data sets (solid lines) and the 
simulations (dashed lines) are shifted vertically relative to each other for 
easier viewing. The circled piece of data set (iii) and the corresponding theory 
are magnified by a factor of 10 for easier viewing. Control field of  400 $\mu$W 
and 3.2 mm diameter (the latter deduced from the Rabi frequency fit) is used. 
The absolute delays are (i) $6$ $\mu$s, (ii) $16$ $\mu$s, and (iii) $59$ $\mu$s, 
yielding delay-bandwidth products (where bandwidth is given by the output pulse) 
of  (i) $0.7$, (ii) $1.2$, and (iii) $2.9$. From the theoretical fit, the 
fraction of energy transmitted is: (i) $18\%$, (ii) $8\%$, and (iii) $0.4\%$. 
\label{fig:slowlight}}
\end{figure}

\textit{Slow light.--} We now consider propagation of a probe pulse 
with intensity envelope FWHM $T = 2.2$ $\mu$s 
in the presence of a CW control field. The fraction of intensity transmitted 
is shown in Fig.\ \ref{fig:slowlight}  with the solid lines 
showing the experiment and the dashed lines showing the excellent theoretical 
fit (see 
below). The three curves, corresponding to different ODs 
increasing from (i) to (iii), exhibit 
large fractional delays that increase 
with OD. 

To model the data, 
we set $\Delta_c = 0$ and determined the Rabi frequency for all three plots in 
the figure by fitting the data in (i) using OD and $\Omega$ as free parameters, 
resulting in OD $= 7.5$ and  $\Omega  = (2\pi) 1.5$ MHz.  
This fit was also used to determine the vertical scale for all three curves, 
because the absolute intensities were not measured. For plots (ii) and (iii) OD 
was used as a free parameter to reproduce the delays, giving an OD of $16$ for 
(ii) and an OD of $59$ for (iii).  Finally, $\gamma_0 = (2 \pi) 5$ kHz was 
adjusted to reproduce the ratio of the maximum intensities of (i) and (ii).  The 
measured fractional delays are very well approximated by $L/(v_g T)$, where the 
group velocity $v_g = L \Omega^2/(\gamma d)$ is given in terms of homogeneously 
broadened optical depth $2 d$ 
\cite{kash99}. The initial short pulse visible on the 
experimental data is $\approx\!1\%$ of the  transmitted off-resonant sideband 
pulse, 
which leaked through the FP etalon after the 
cell.

In summary, we combined buffer-gas cooling with laser ablation to obtain a new and simple method to generate novel coherent optical media. 
As an example of this method, we cooled $^{87}$Rb by He buffer gas to below 7 K to produce an optically thick medium with  low ground state decoherence, excellent EIT, strong enhancement of nonlinearities, and large fractional delays in slow light propagation.
By optimizing such factors as buffer gas pressure, cell geometry, magnetic shielding, and the ablation procedure, we expect to increase OD, decrease 
decoherence, and tailor the system to specific applications. Our new system 
combines the slow atomic motion and good coherence properties of MOT systems 
with the simplicity and large atomic volume of vapor cell systems. Moreover, our system can be used to produce coherent optical media of other atoms, molecules, or their combinations by simply replacing the source. All these features distinguish our method and this particular implementation of it as promising candidates for the use in nonlinear and quantum optics as well as in precision measurements. In particular, applications to delay lines, 
quantum memories, and sources of single or paired photons can be foreseen. 
 
We thank E.N. Fortson for lending us a diode laser, A.A. Zibrov for help in 
building the setup, and M. Parsons, W. Campbell, E. Tsikata, M. Hummon, and M. 
Hohensee for 
discussions. This work was supported by NSF CUA and GRFP, DARPA, NSF 
grant PHY-0457047. 


\begin{thebibliography}{99}
\bibitem{odnote} 
Due to the variation of OD from one ablation to another, 
OD is used below as a fit parameter. 
\end{thebibliography}

\end{document}